\documentclass[preprint,nofootinbib,showpacs,aps]{revtex4}
\def\be{\begin{equation}}
\def\ee{\end{equation}}
\begin{document}
\title{Reply to the Comment ``On the thermodynamics of inhomogeneous 
perfect fluid mixtures''}
\author{Hernando Quevedo}\email{quevedo@physics.ucdavis.edu}
\affiliation{Instituto de Ciencias Nucleares\\
Universidad Nacional Aut\'onoma de M\'exico\\
 P.O. Box 70-543, M\'exico D.F. 04510\\}
\affiliation{Department of Physics\\ University of California \\
Davis, CA 95616}
\author{Rub\'en D. Z\'arate}
\email{zarate@nuclecu.unam.mx}
\affiliation{ Instituto de Ciencias Nucleares\\
     Universidad Nacional Aut\'onoma de M\'exico \\
     P.O. Box 70-543, M\'exico D.F. 04510}

\begin{abstract}
We show that the analysis presented in a recent comment by 
Coll and Ferrando \cite{comment}  (qr-qc/0312058)
is based on the erroneous assumption that the 
chemical potential and fractional concentration of a {\it mixture}
of perfect fluids are unknown variables.
\end{abstract}

 \noindent \noindent 

\maketitle

In a previous comment \cite{comment} on a recent paper by the 
present authors \cite{zq}, Coll and Ferrando argue that
the thermodynamic scheme for a binary mixture of perfect fluids
is trivially satisfied by any perfect fluid solution 
of Einstein's equations. We will show here that this 
conclusion is erroneous due the fact that when one assumes
the existence of a perfect fluid binary mixture, one can not
consider the chemical potential and the fractional concentration
as unknown variables.

Let us first mention that the comment \cite{comment} can be divided into two
parts. The first and essential part is contained in proposition
1 which asserts that any perfect fluid solution is compatible
with the binary thermodynamic scheme proposed in \cite{zq}. 
The second part contains
a detailed analysis of proposition 1 and illustrates its 
consequences in the case of the Szekeres and Stephani universes.
Consequently, it is enough to concentrate on proposition 1.

The proof of proposition 1 is based on Eqs.(4) and (5) of \cite{comment}. 
In Eq.(4), Coll and Ferrando consider the 1-form
\be
\omega = {\rm d}(\rho/n) + p {\rm d} (1/n) 
\label{badgibbs}
\ee
in a 4-dimensional space, and  
argue the there exist functions $s$, $c$, $T$, and $\mu$ such that
the 1-form (\ref{badgibbs}) can be written as 
$\omega = T {\rm d} s + \mu {\rm d} c$. 
This is obviously true
as a consequence of Frobenius theorem. (The use of the Pfaffian
method is not completely appropriate in this case, as stated in 
\cite{comment}, because 
$\omega$ is not a Pfaffian form.)  
The question is whether
the 1-form (\ref{badgibbs}) corresponds to the Gibbs 1-form
of a {\it binary mixture} of perfect fluids and whether it
can be used to analyze the thermodynamics of such a system.
The answer is negative. Indeed, if we assume that we have
a solution of Einstein's equations which represents a 
{\it binary mixture} of perfect fluids, it means that 
we know the thermodynamic variables corresponding to 
the total energy-density $\rho$, the pressure $p$, the total 
particle number density $n$, and also the {\it composition 
of the mixture}, i.e,  the chemical potential $\mu$
and the concentration $c$. If the last two thermodynamic
variables are not known, it is impossible to establish
if we are dealing with a single perfect fluid or with
a mixture of perfect fluids. This implies that the 
statement in proposition 1 of \cite{comment} is not 
correct. 

Consequently, the correct Gibbs 1-form for a binary mixture is 
\be 
\Omega = 
{\rm d}(\rho/n) + p {\rm d} (1/n) - \mu {\rm d} c
\label{rightgibbs}
\ee
and the correct question to be asked is whether 
there exist functions $T$ and $s$ such that 
$\Omega = T{\rm d} s$. In this case, the Frobenius 
condition is not identically satisfied. Indeed, Frobenius
theorem asserts that $\Omega \wedge {\rm d} \Omega = 0$
is a necessary and sufficient condition for the existence
of the functions $T$ and $s$. The chemical potential 
and the fractional concentration are {\it known} functions
and for this reason they can be written in terms of the
known variables $\rho$, $p$, and $n$ as: 
$\mu = (\rho + p)/n$ and $c=\ln(c_0/n)$ [cf. eqs.(27) and
(29) in \cite{zq}]. 
The analysis performed 
in \cite{zq} for the Szekeres and Stephani universes show 
that the condition 
$\Omega \wedge {\rm d} \Omega = 0$
is, in general, not compatible with the 
entropy production condition $\dot s = - (\mu/T) \dot c $.

Thus, we have shown that it is not possible to use the 1-form
(\ref{badgibbs}) to investigate the thermodynamic properties
of a binary mixture. But if we insist in using it, we certainly
will find the set of functions $T$, $s$, $\mu$, and $c$ for the
new representation of $\omega$. However, one would see that 
these functions can not be interpreted as temperature, entropy
density, chemical potential and fractional concentration, respectively.
This would be an additional example of a well-behaved solution to a 
well-posed mathematical problem which, nevertheless, has no physical
significance.

\section*{Acknowledgements}
  This work was supported  by DGAPA-UNAM grant IN112401,  
CONACyT-Mexico grant 36581-E, and US DOE grant DE-FG03-91ER 40674.
R.D.Z. was supported by a CONACyT Graduate Fellowship.
H.Q. thanks UC MEXUS CONACyT (Sabbatical Fellowship Programm) 
for support.


\begin{thebibliography}{99}


\bibitem{comment} Coll B and Ferrando J J 2004 Comment on the thermodynamics
of inhomogeneous perfect fluid mixtures, Class. Quantum Grav. (2004)
submitted. {\it Preprint} arXiv:qr-qc/0312058


\bibitem{zq} Zarate R D and Quevedo H 2004 {\it Class. Quantum Grav.}
{\bf 21} 197 [arXiv:gr-qc/0310087]



\end{thebibliography}
\end{document}